\documentclass{article}

\usepackage[preprint]{neurips_2023}

\usepackage{caption}
\usepackage{float}
\usepackage{multirow}
\usepackage[utf8]{inputenc} 
\usepackage[T1]{fontenc}    
\usepackage{hyperref}       
\usepackage{url}            
\usepackage{booktabs}       
\usepackage{amsfonts}       
\usepackage{nicefrac}       
\usepackage{microtype}      
\usepackage{xcolor}         
\usepackage{gensymb}

\usepackage{graphicx}
\graphicspath{ {./images/} }

\title{Machine Learning–Based Prediction of Heat Index in Selected U.S. Cities}

\begin{document}

\author{
  Yushan Han\footnotemark[1]\\
  Department of Land, Air and Water Resources \\
  University of California, Davis \\
  \texttt{yshhan@ucdavis.edu} \\
  \And
  Calen Randall\footnotemark[1] \\
  Department of Land, Air and Water Resources\\
  University of California, Davis\\
  \texttt{cfrandall@ucdavis.edu} \\
}

\footnotetext[1]{The two authors contributed equally to this work. Yushan Han contributed to conceptualization, methodology, model training, writing, and visualization. Calen Randall contributed to conceptualization, data processing and analysis, writing, and visualization.}

\maketitle

\begin{abstract}
Heat stress has harmful effects that impact communities across the Unitedt States, particularly when high temperatures are accompanied by high humidity. The combined impact of temperature and humidity can be summarized by the heat index (HI). Current state-of-the-art numerical weather prediction models are often biased when forecasting temperature and humidity even within a 24-hour forecast lead time. This study explores the ability of machine learning (ML) models to accurately predict the next-day heat index using Random Forest and single-layer Gated Recurrent Unit (GRU) models in four locations across the United States. We find that Random Forest and GRU models perform reasonably well at all four selected locations. Mean absolute HI error ranges from 4.5 to 6.6\degree F. All model versions have an accuracy rate exceeding 80\% in three of the four locations in terms of successfully forecasting an extreme heat day, as indicated by a high afternoon HI. The GRU model achieves over 95\% accuracy in these three locations. Model performance details vary by location. In Minneapolis and Portland, which have relatively few days with high HI values, models' accuracy is high, but the recall and precision are generally very low. In contrast, Dallas, a location with many high HI days, shows moderately high accuracy, as well as extremely high recall and precision. These differences are likely due to distinct causes of heatwaves in different climatological regions of the United States, as reflected in the feature importance scores output by Random Forest models. The ML models designed in this study can be used to assist with local heat index forecasting and extreme heat warning issuance at minimal computational cost.
\end{abstract}

\section{Introduction}

Extreme heat events are one of the most prominent and damaging climate extremes. They cause significant health stress impacts on humans and ecosystems, damage infrastructure, destroy agriculture, and can exacerbate other extreme weather events such as wildfires, drought, and dust storms. In the USA, they are the leading cause of death for weather-related fatalities and are responsible for thousands of hospitalizations annually (EPA 2022). Given their harmful impacts on human health, the atmospheric science community devotes significant focus to the forecasting of extreme heat events and modeling of how extreme heat events are expected to change in the future. With increasing average temperatures due to climate change, the probability of extreme heat events exceeding past thresholds drastically increases (Dahl et al., 2019, EPA 2022, Perkins et al., 2012), necessitating more heat-prepared communities and more skillful heatwave forecasting techniques.

To properly assess the impact of extreme heat on human health, more comprehensive variables than raw temperatures must be considered. For example, a person might feel more comfortable during heat events where the humidity is low compared to a heat event of similar temperature with higher humidity. When the atmosphere is humid, sweat cannot evaporate, leaving a person more vulnerable to heat stress. Sensitive populations, people who work outdoors, and people living in buildings without heat resilient infrastructure are particularly vulnerable to heat stress during these conditions.

The Steadman Heat Index (Steadman, 1979) is a metric that combines temperature and humidity to calculate a "feels like" temperature, which is useful for quantifying the risk of heat stress. While the original Steadman heat index calculation includes comprehensive parameters such as "effective radiation area of skin" (the surface area of a person that can effectively perspire) and "activity" (the metabolic output or how intensely a person is working), the National Weather Service (NWS) has simplified it into a regression equation that uses more common atmospheric variables such as air temperature and relative humidity of an air mass (Rothfusz, 1990). Based on this algorithm, the NWS has developed a chart of NWS Heat Index broken into four categories: Caution (above 80\degree F), Extreme Caution (Above 90\degree F), Danger (above 103\degree F), and Extreme Danger (above 125\degree F). Heat index in the Danger and Extreme Danger categories are rare, except for locations in the US southeast. From this chart we can see that the heat index is positively correlated with temperature and relative humidity when the air temperature is above 80\degree F.

The two variables on which the heat index depends---near-surface (2-meter) temperature and relative humidity---are often subject to considerable bias in forecasts produced by conventional weather models, even within a 24-hour lead time. This is primarily due to  complexities in terrain and vegetation patterns, the simplification and parameterization processes used to model the Earth's surface and land-air interaction, and the omission of processes such as irrigation in the models, which result in an inaccurate representation of soil moisture and surface heat fluxes (Yu et al., 2025). In the Indian monsoonal region, research (Kantha Rao and Rakesh, 2019) found that the 2-meter temperature produced by the Weather Research and Forecasting (WRF) model can have over 15\% relative error in some regions within a 24-hour forecast period. The projected 2-meter relative humidity can also be overestimated by over 10\% in some cases, particularly during the early morning and mid-afternoon when minimum and maximum temperatures occur. Alexander et al. (2022) also showed that daily maximum 2-meter temperature at the California Central Valley (CCV)---a region featuring regular irrigation and complex surrounding terrain---can have a warm bias of about 4 \degree C (or 13\% relative error) in high-resolution WRF model. The surface moisture within CCV in some model simulations was off by nearly 50\%. These biases, when translated to the heat index, can be as high as 10°F or even more. The large uncertainties in these model forecasts jeopardize local meteorological agency's ability to accurately issue heat alerts to the public.

Heterogeneity in the topography and climatology of the United States leads to characteristics of extreme heat to vary regionally. Mechanisms driving extreme heat events are not homogeneous across the USA. Extreme temperatures are often driven by a corresponding high pressure ridge (often referred to as persistent or blocking highs), however, in some regions extreme temperatures occur when the high is located adjacent rather than directly above, enhancing the temperature through horizontal temperature advection. Extreme heat is highly tied to the shape of the local June-July-August (JJA) temperature distribution. Regions featuring more positively skewed JJA temperature distributions (such as the Pacific Northwest and Pacific Coast) tend to experience infrequent but intense extreme temperatures, while in regions of negatively skewed JJA temperature distributions (the Desert Southwest, Southeastern USA) extreme heat is more tempered relative to seasonally average JJA temperatures. However, the enhanced humidity during periods of extreme heat drives high heat index values throughout these regions. The mechanisms of moisture-induced extreme heat are influenced by soil moisture regimes (Benson and Dirmeyer, 2021). Tthe eastern USA is more humid during JJA and is less sensitive to the presence of high pressure systems leading to extreme temperatures. In contrast, The western USA is generally soil moisture limited during JJA and thus is more sensitive to soil moisture decreases leading to extreme temperatures.

Recent developments in machine learning (ML) methods have made it possible for ML-based models to compensate for the limitations of current weather models in predicting the next-day heat index or heat index category at a given location. These models can also assist local meteorological agencies in determining whether to issue alerts to the public based on heat index categories. ML models can also provide insights into which factors play the most vital role in modulating heat events in different regions. Deep learning (DL) architectures, such as spatial-temporal transformers and convolutional neural networks, have been widely used to simulate three-dimensional atmospheric variables globally, just like many global numerical models. DL-based models based on these architectures have achieved superior accuracy in forecasting atmospheric variables at large scales (e.g., Bi et al., 2023; Lam et al., 2023). However, these models are not designed and optimized to predict the heat index at local scales (Hill et al. 2020).

\begin{figure}
    \centering
    \includegraphics[scale=0.4]{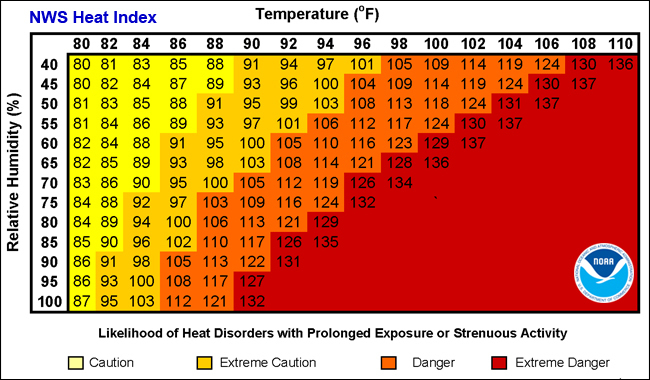}
    \caption{NWS Heat Index displaying the Temperature and Relative Humidity criterion for each category of heat stress likelihood.}
\end{figure}

Simpler ML models have been developed for specific tasks like predicting daily average temperature, humidity, and rain or storm events using an array of atmospheric input data. While random forest are better at predicting categorical and probabilistic forecasts such as the probability of severe weather in the coming days (Gad and Hosahalli, 2022; Hill et al., 2020), artificial neural networks (ANN) algorithms that utilize weighted node layers including input, output, and hidden layers are more often used to predict continuous variables like temperature and humidity, and they are proved to produce higher accuracy results compared to basic machine learning algorithms such as random forest (Astsatryan et al., 2021; Hanoon et al., 2021). Very few studies focus specifically on predicting the heat index. Tumse et al. (2023) produced a monthly heat index map of Türkiye using an ANN. To the authors' knowledge, no rigorous research has studied the ability of ML models to justify heat stress risks based on projected afternoon heat index and heat index category at the local scale.

The main objective is to train both categorical and regression models to effectively predict whether the next day's heat index will exceed the NWS-defined threshold of 80\degree or 90\degree F (Caution or Extreme Caution) based on an array of atmospheric variables in a 1-day (>24-hour) lead time in the four selected cities, each of which represents a different climatological and topographical region in the United States. While model accuracy is desirable, this paper primarily focuses on comparing and illustrating the results of different ML models and different forecast locations.

\section{Data}

Data for the project comes from the European Center for Medium-Range Weather Forecast’s (ECMWF) ERA5 reanalysis data. ERA5 reanalysis data combines observations and modeled weather data interpolated onto a latitude-longitude grid on quarter degree horizontal resolution and vertical resolution based on constant levels of atmospheric pressure (units of hectopascals or millibars). Data is saved at hourly intervals and spans from 1979 to 2025, though data for this project was selected for the years 1979-2022. Predictor variables were selected from 10:00 a.m local time at four vertical levels: surface, the 850 hPa pressure level (approximately 1500m above sea surface level, typically within the well-mixed planetary boundary layer), 700 hPa (approximately 3000m), and 500 hPa (provides synoptic level features, approximately 5600m). Data was selected from the months May-September as large heat index values outside of this time period are rare in the USA. 

The target variable, Heat Index (HI) at 4 p.m., which is the most common time of day when the near-surface temperature peaks during summer time, is calculated daily using the NWS equation that approximates the HI regression equation in Rothfusz (1990) for simplicity: 

\[HI = 0.5 * {T + 61.0 + [(T-68.0)*1.2] + (RH*0.094)}\]

where T in our data is the 4 p.m. surface temperature (in Fahrenheit) and RH is the surface relative humidity at the same time. To ensure HI calculation accuracy, the original regression equation is used for heat index values greater than 80\degree F:

\[HI = -42.38 + 2.05*T + 10.14*RH - 0.22*T*RH - 6.8*10^-3*T^2 - 0.055*RH^2 \]
\[+ .0012*T^2*RH + 8.5*10^-4*T*RH^2 - 1.9*10^-6*T^2*RH^2\]

At the surface level, temperature, dew point, and wind direction are selected as predictor variables. At the 850 hPa level, temperature, subsidence, and RH are selected. At 700 hPa, subsidence and RH are chosen. Finally, at the 500 hPa level, geopotential height and subsidence are selected as predictor variables. Geopotential height is related to the temperature in the vertical direction at a specific geographic point and gives context to the general synoptic pattern, specifically whether there is a high pressure “ridge” (higher geopotential) that acts as a driver of anomalously warm weather patterns. Surface and low-level wind speed and direction were included for additional weather pattern and “direction-of-flow” context. In some regions, extreme heat is driven by the direction of airflow. For example, across the California Central Valley and coast, heatwaves are associated with offshore flow (winds blowing from the north or east) which brings warms and dry continental air (Clemesha et al., 2018, Grotjahn and Faure, 2008). Westerly or southerly winds are associated with cool and moist air masses from the Pacific Ocean which inhibit temperature extremes. Subsidence, the measure of average vertical motion in the atmosphere, corresponds to the general weather pattern. In a pattern with high geopotential and high surface pressure, subsidence is likely negative, as warm dry air travels down towards the surface. Highly negative subsidence values are generally required during periods of extreme heat so that convection, stemming from the warm surface, is inhibited. Additionally, the sum of precipitation between 3 a.m. and 9 a.m. was also included. Because all variables are recorded at 10:00 am one day before the forecast day, our goal of predicting HI at least 24 hours in advance is met. The final variables included in our analysis are listed in Table 1. 

\begin{table}[H]
\centering
\begin{tabular}{|c c c|} 
 \hline
 Variable Name & Level (hPa) & Units \\ 
 \hline
 Temperature & Surface, 850, 700 & Kelvin/Celsius \\ [1ex]

 Dew Point & Surface & Kelvin/Celsius \\[1ex] 

 Relative Humidity & 850, 700 & Unitless \\[1ex] 

 Wind Speed & Surface, 850 & Meters per Second \\[1ex] 

 Wind Direction & Surface, 850 & Degrees \\ [1ex] 

 Geopotential & 500 & Hectopascals \\ [1ex] 

 Subsidence & 850, 700 & Meters Per Second \\ [1ex] 

 3 a.m.-9 a.m. Cumulative Precipitation & Surface & Millimeters \\ [1ex] 
 \hline
\end{tabular}
\\ [1ex]
\caption{A list of predictor variables used to train the models in this study.}
\end{table}

Four cities from four regions of distinct heatwave characteristics were selected (Pacific Northwest, Great Plain, Gulf Plains, and Northeast respectively): Portland (45.5\degree N -122.75\degree W), Minneapolis (45.0\degree N,-93.25\degree W), Dallas (32.5\degree N, -97.0\degree W), and Boston (42.25\degree N, -71.25\degree W). Predictor variable data at these four locations in ERA5 are collected for analysis and model training for each individual city.

Different heat index thresholds were defined for each city as the distribution of heat index values varies greatly depending on the location (Figure 2). In Dallas, 64.9\% of all days from May-September have a heat index greater than 90\degree F in contrast with a mere 4.1\% of Portland days that break the 90\degree F threshold. A threshold of 80\degree F, NWS categorized "Caution" was chosen for Portland (26.4\% above 80\degree F) and Boston (32.4\% above 80\degree F). The NWS defined "Extreme Caution" 90\degree F threshold was selected for Dallas and Minneapolis (8.2\% above 90\degree F). 

\begin{figure}[H]
\includegraphics[scale=0.33]{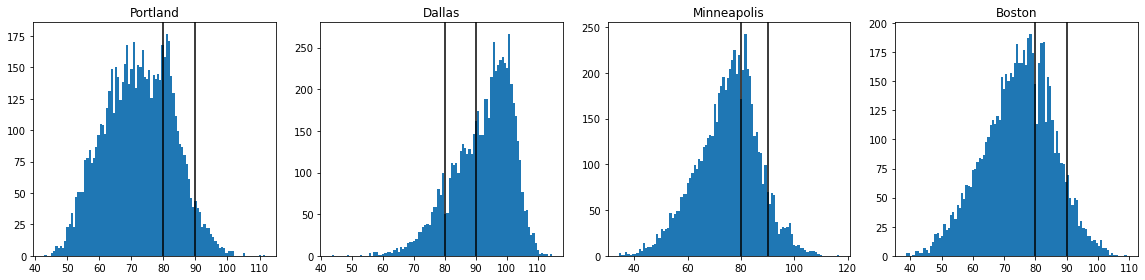}
\caption{Distribution of Heat Index for all four cities}
\end{figure}

A common data issue when running machine learning models on weather data is the analysis of wind direction data which is defined via the unit of degrees. Wind directions of both 5 and 355 degrees are classified as "Northerly", however the model treats the two values as on opposite ends of the wind direction spectrum. There are several techniques to avoid such distinction. These solutions were deemed unnecessary following analysis of wind directions at all four cities using wind roses, which plot the frequency of binned wind speeds for specified wind directions. As shown in Figure 3, very little overlap occurred between northerly winds with direction values greater than 345\degree and less than 15\degree for all four cities. Northerly winds either had a more westerly or more easterly component. The wind rose results for days that satisfy the defined threshold, above 80\degree F for Boston and Portland and above 90\degree F for Minneapolis and Dallas, had similarly little northerly wind overlap, except for Portland. Days with high heat index occurred on days with primarily southwesterly winds in Dallas and Minneapolis, westerly winds in Boston, and easterly winds in Portland, though in Portland, 18.4\% of the winds fell within 0-15 degrees and 9.1\% of the winds were greater than 345 degrees, so some northerly wind overlap occurred. In the other three cities, northerly winds accounted for no more than 2\% of the high heat index days.

\begin{figure}[H]
\includegraphics[scale=0.45]{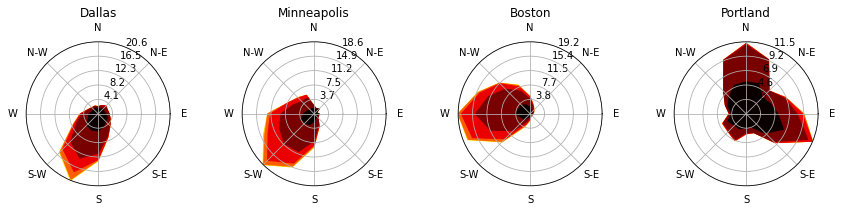}
\caption{ Wind Rose for days when the next day's Heat Index is above 90 F}
\end{figure}

\section{Methods}

To accomplish the above tasks, we leverage random forest models and a single-layer Gated Recurrent Unit (GRU) with different input data time steps. We then compare the performance of different models. We choose random forest algorithms because they are suitable for producing categorical forecasts (i.e., whether the heat index meets a threshold) and for ranking the importance of predictor variables in causing hot days. GRU is a type of improved Recurrent Neural Network (RNN) with a gating mechanism (Chung et al., 2014). Compared to the vanilla RNN model, GRU better handles the vanishing gradient and performs faster. Like all RNNs, GRU is adept at maintaining a sizable memory, allowing for the input of lengthy time series data. Although GRU is frequently employed with longer time steps to enhance its capacity for holding memory, GRU with one time step may also be advantageous when the goal is a single-step forecast and the inputs do not exhibit significant temporal patterns or long-term dependencies. In this case, GRU is similar to a feedforward neural network, however, it has a more advanced architecture that controls the data flow via its gates, which can extract more complex patterns of data within just one time step. 

As Schultz et al. (2021) pointed out, although ML-based weather models can efficiently handle large-scale data and produce better forecasts in most cases, they are not good at predicting extreme weather events because there is little precedent for the models to learn from. In this study, we set up a challenging situation for machine learning models to predict statistically extreme HI in the case of Minneapolis, where days with over 90\degree F HI are rare.

\begin{figure}[H]
\centering
\includegraphics[scale=0.6]{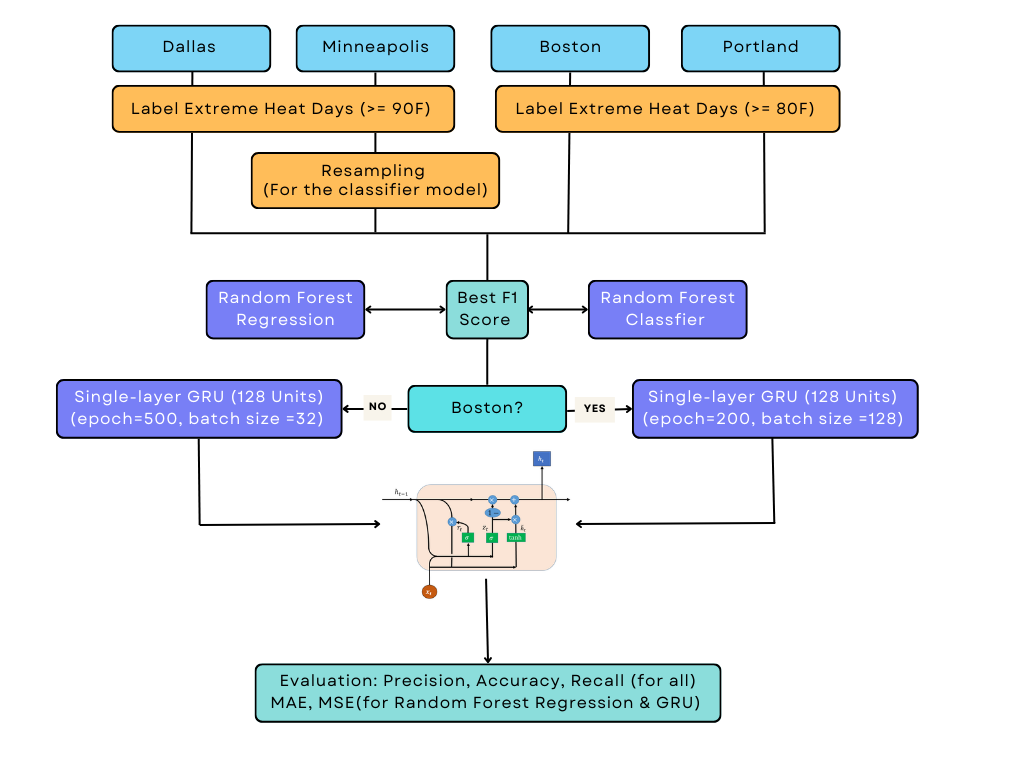}
\caption{The workflow of our methods }
\end{figure}

\subsection{Random Forest}

In this study, we use two random forest models: the random forest regression model and the random forest classifier model. If one day is followed by an extreme heat day---a day with a heat index above the heat wave threshold of 80 or 90\degree F---that day is labeled "1" in the datasets and forecast outputs. Otherwise, it is labeled "0." Since Minneapolis has only 8.2\% extreme heat days over 90°F, we balance the input data for the classifier model by oversampling days that meet the threshold to reach a 1:1 ratio of 1s to 0s. The resampling process is designed to improve the model's performance in this unbalanced case.

The inputs are 13 meteorological predictors and the outputs are the next-day extreme heat day label for the classifier model and the HI value for the regression model. The input data was split in half. The first half (1979–2000) represents the historical climate data and training data, while the second half (2000–2022) represents the validation data. This split allows us to test the effectiveness of using 20th-century weather data to forecast 21st-century heat waves under ongoing climate change.

\subsection{GRU}

For the GRU model, the training and validation datasets are split in half, just like we did for random forest models. For the purpose of comparing the performance of deep learning GRU models to that of random forest models, the input data time step is initially set to be one (one day's meteorological variables) in order to match the input data dimension for random forest models. The output data is an exact HI value. We build a 128-unit single-layer GRU with the default "tanh" activation function.  Before feeding data into the models, we scale all training input data to the range of 0 to 1. We set the batch size to the default of 32 and the epoch number to 500. Except for the Boston case, we find that the model is very likely to overfit. Therefore, we increase the batch size to 128 and decrease the epoch number to 200, which improves the model's performance for the Boston case. The output layer is a one-unit dense layer containing a single-step HI value.

We try increasing the time step to take advantage of the GRU's memory-holding ability. However, doing so degrades the model performance for the validation dataset, though it improves the performance based on the training data. See Section 4 for details. This is somewhat expected, as the heat index on a single day is not that relevant to meteorological factors several days ago. Longer time steps are better suited for predicting long-term temperature trends.

A summary of the workflow of our method is illustrated in Figure 4.

\subsection{Evaluation Metrics}

 For both models of random forest, we tune the model's tree depth parameter to output the best forecast for HI. The forecast skill is evaluated by the F1 Score, which is the harmonic mean of the precision [true positive/(true positive+false positive)] and recall [true positive/(true positive+false negative]. We tested a range of different tree depths for both models and selected the optimized tree depth from the trained model with the highest F1 score. Random forest has the benefit of highlighting the significance of each variable influencing the model's outcomes. Hence, we also evaluate the importance of each meteorological predictor using the mean decrease in impurity. The measures of the GRU model performance uses mean absolute error (MAE) and mean squared error (MSE) of predicted HI.

 To evaluate the performance of the models in successfully forecasting extreme heat days when the Heat Index (HI) exceeds alarming thresholds, extreme heat day labels are assigned to the regression models. If the predicted HI exceeds the prescribed threshold, a "1" label is assigned to that day for that model. All models are evaluated based on the labels using accuracy [(true positive + true negative)/total predictions], precision, and recall.
 
\section{Results}

\begin{figure}[H]
\centering
\includegraphics[scale=0.37]{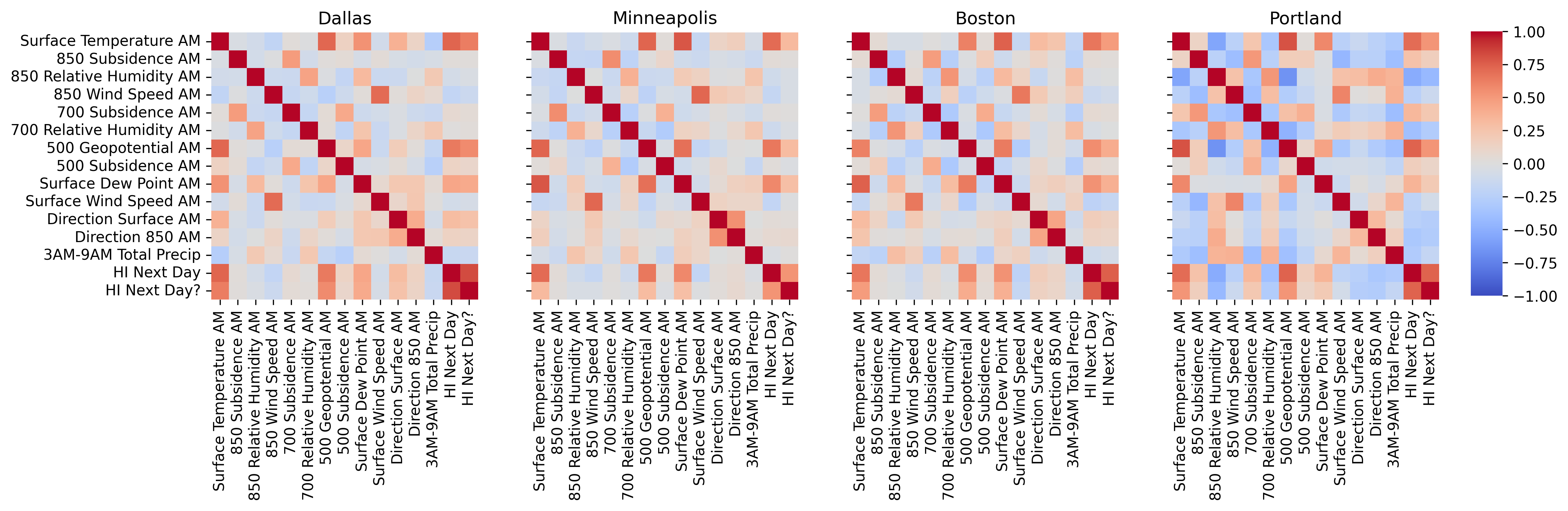}
\caption{Heat map for cross-variable correlation coefficient for the four cities}
\end{figure}

First, we investigate the cross-variable correlations as shown in Fig. 5. Since we already discarded highly correlated variables in the initial dataset, the heat map shows no outstandingly strong correlation coefficient among different meteorological variables. As one might expect, the next-day heat index is moderately positively correlated with surface temperature, 500 hPa geopotential height (representing the mean temperature of the entire atmospheric layer, or the synoptic weather pattern), surface dew point (representing surface humidity level), and subsidence rate. The heat index is most often negatively correlated with wind speeds and early morning total precipitation, while the relationship between the heat index and wind direction varies from city to city.

\subsection{Model Results}

\begin{figure}[H]
\centering
\includegraphics[scale=0.45]{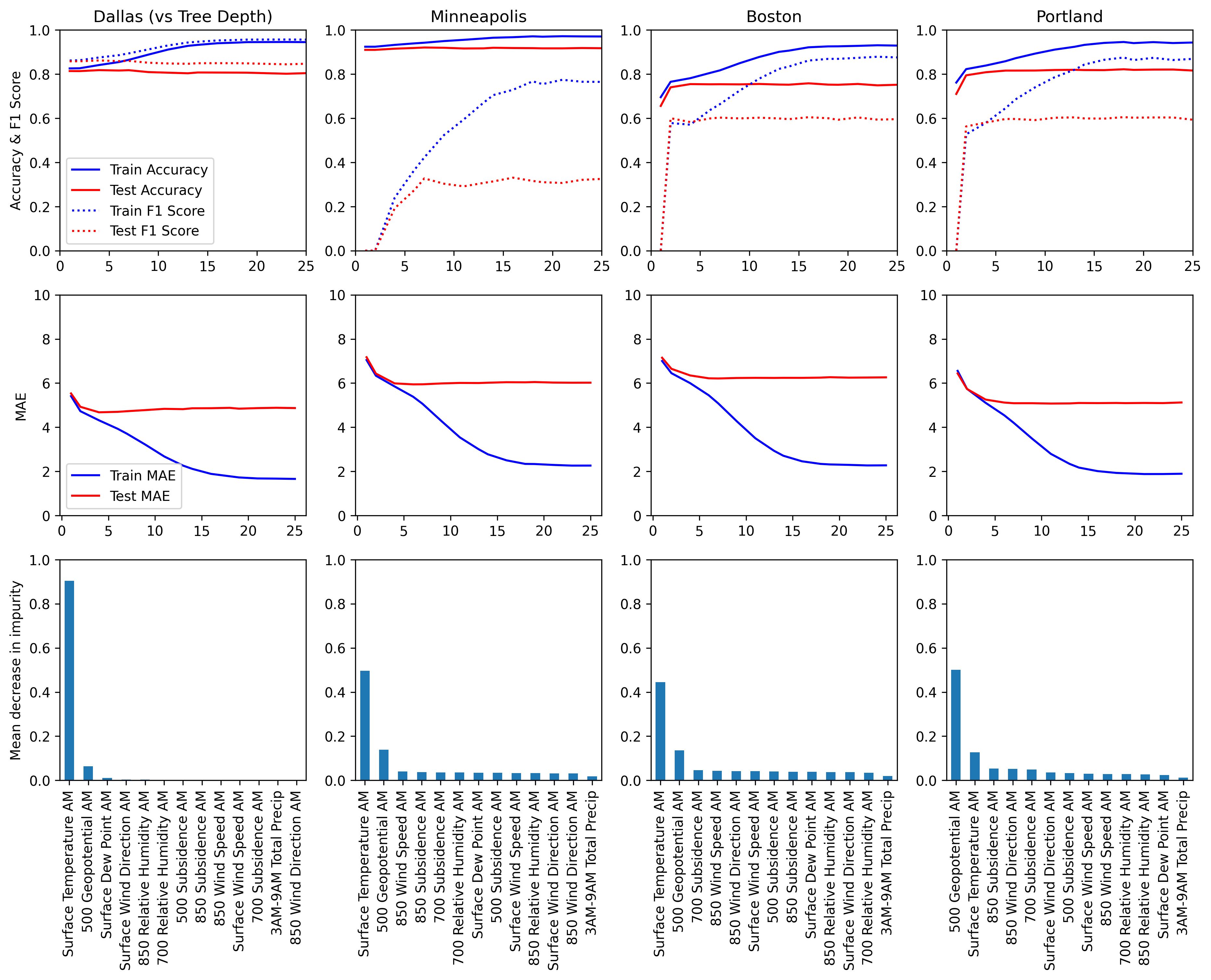}
\caption{Random forest regression model results}
\end{figure}

As shown in Fig. 6, it is typical to have the validation F1 score and MAE trailing the training F1 score and MAE, and the difference between the two can be large no matter how we tune up the parameters for the regression random forest model. Thus, the model validation performance saturates at a small tree depth, indicating overfitting. It might be inferred that the heatwave pattern is changing over time as global warming accelerates and the frequency of extreme heat days increases. This could also imply an insufficient amount of information for building a more precise model to validate future predictions. The subfigures of importance in the last row of Fig. 6 are calculated based on the fitted model with the optimized tree depth, which is evaluated by the highest F1 score. The figure clearly shows the dominant factors, such as morning surface temperature, especially for Dallas. The 500 hPa geopotential is usually the second most significant contributor to the model, except for Portland. Portland has a more evenly distributed contribution from the variables, and the wind direction shows some significance as well.

\begin{figure}[H]
\centering
\includegraphics[scale=0.55]{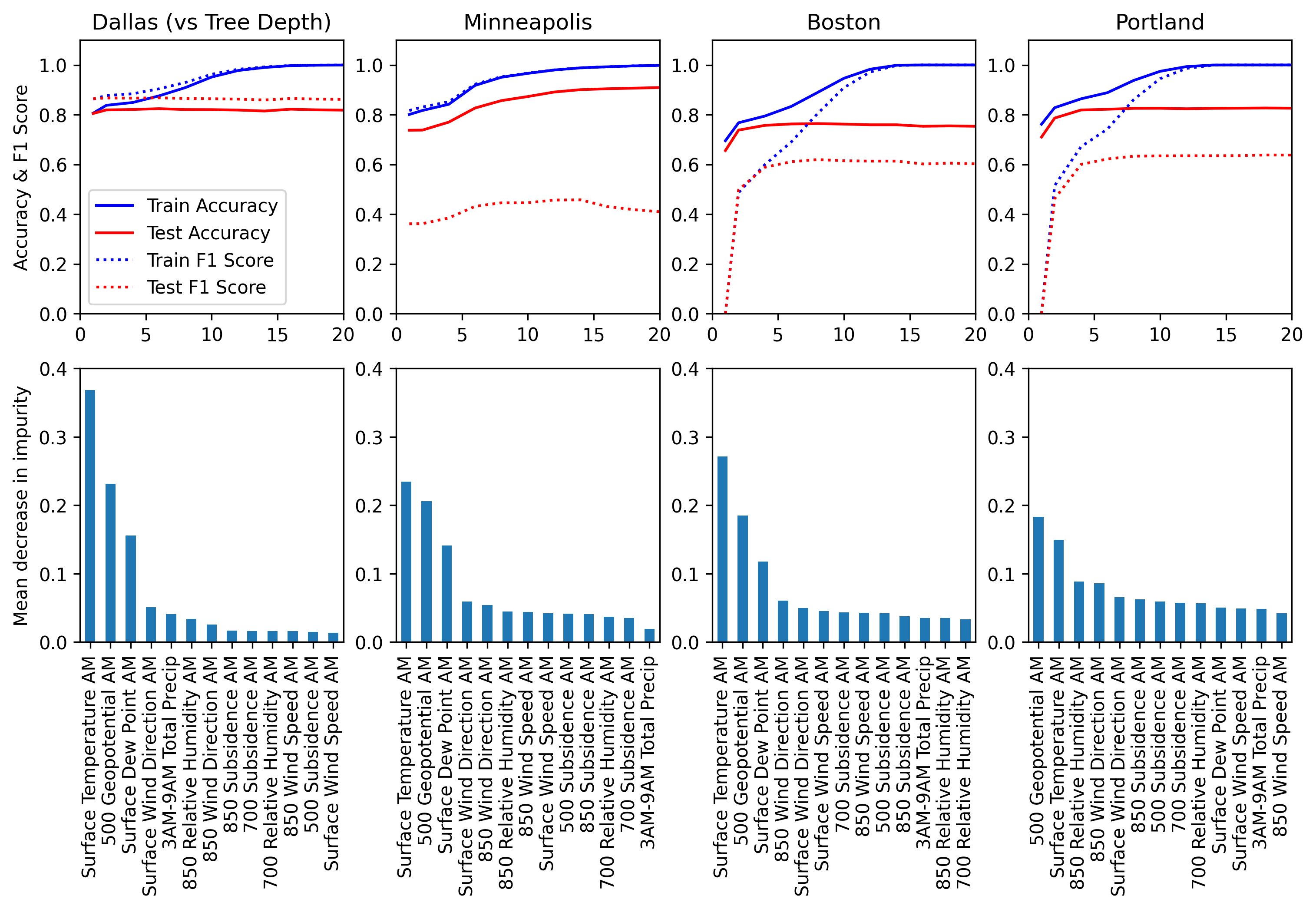}
\caption{Random forest classifier model results}
\end{figure}

\begin{figure}[H]
\centering
\includegraphics[scale=0.45]{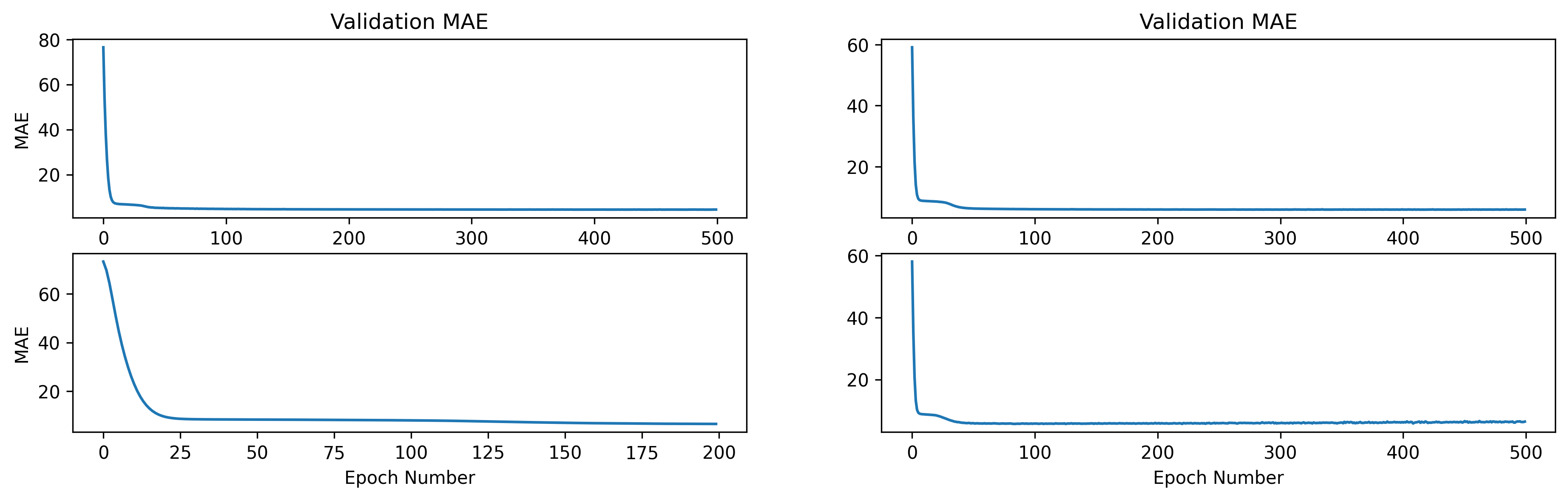}
\caption{Single-layer GRU model results. Dallas (Top left), Minneapolis (Top right), Boston (Bottom Left), Portland (Bottom right)}
\end{figure}

Figure 7 shows the performance of the random forest classifier. Similar to the random forest regression model, the classifier's validation suffers from small deductions in the loss function and small gains in the reward at greater tree depth. However, the gradient of the reward function (F1 score) is slightly smaller than that of the regression model. The importance of the variables also demonstrates this difference. The importance of the different variables for the four cities is much more comparable than in the regression model. As in the previous model, Portland's significant variables are distinctive. The importance of the variables in different cities is discussed in more detail in Section 5. The one time step single-layer GRU model's epoch time evolution is shown in Fig. 8. The results show that the loss function quickly decreases for all cities where the model converges. No overfitting is observed.

\subsection{Comparisons}

\begin{table}[H]
\centering
\begin{tabular}{l l l c c c}
\hline
\textbf{Model} & \textbf{City} & \textbf{MAE (MSE)} & \textbf{ACC} & \textbf{REC (PRE)} \\
\hline

\multirow{4}{*}{Random Forest Regression}
& Dallas, TX        & 4.67 (39.58) & 82\% & 85\% (87\%) \\
& Minneapolis, MN  & 6.04 (60.94) & 92\% & 21\% (62\%) \\
& Boston, MA       & 6.25 (62.14) & 76\% & 55\% (68\%) \\
& Portland, OR     & 5.09 (40.58) & 82\% & 47\% (83\%) \\
\hline

\multirow{4}{*}{Random Forest Classifier}
& Dallas, TX        & - & 82\% & 87\% (86\%) \\
& Minneapolis, MN  & - & 90\% & 47\% (45\%) \\
& Boston, MA       & - & 76\% & 56\% (70\%) \\
& Portland, OR     & - & 83\% & 53\% (81\%) \\
\hline

\multirow{4}{*}{GRU Model (timestep=1)}
& Dallas, TX        & 4.45 (39.50) & 97\% & 99\% (61\%) \\
& Minneapolis, MN  & 5.90 (60.50) & 96\% & 61\% (100\%) \\
& Boston, MA      & 6.58 (69.64) & 86\% & 100\% (37\%) \\
& Portland, OR     & 6.40 (60.61) & 98\% & 99\% (73\%) \\
\hline

\multirow{4}{*}{GRU Model (timestep=2)}
& Dallas, TX      & 4.52 (40.44) & 85\% & 93\% (86\%) \\
& Minneapolis, MN & 5.98 (62.33) & 91\% & 30\% (48\%) \\
& Boston, MA      & 6.51 (68.07) & 72\% & 78\% (20\%) \\
& Portland, OR    & 5.90 (52.54) & 93\% & 47\% (31\%) \\
\hline \\[1ex]
\end{tabular}
\caption{Model performance summary. REC stands for recall, PRE stands for precision, and ACC stands for accuracy}
\end{table}

Table 2 provides a summary of the model performance results, which are evaluated using different metrics. Minneapolis is highlighted because we are interested in observing how the models perform when predicting rare incidences, such as the greater-than-90-day incidence in Minneapolis. Although the overall accuracy is similar, given that the "0" class is easy to predict, recall (hit rate) and precision differ. The difference between recall and precision is significant for all models except the classifier. Resampling the input data for the random forest classifier model helps balance recall and precision, though it results in a lower F1 score than the regression model counterpart, as shown in Figures 6 and 7. The one-time-step GRU model performed best for the challenging Minneapolis case. 

Overall, the performance of the random forest regression and the classifier models are comparable, with the GRU model performing slightly better in terms of recall and accuracy. However, the GRU model may have a higher mean squared error (MSE) than the other models. For example, this occurs in the cases of Boston and Portland. Examining the predicted data for these two cases, we find that the model underestimates the hottest days and overestimates the coolest days. Additionally, when we implemented the two-step GRU model, its performance degraded compared to the one-step model in all four cities. Therefore, we stopped attempting to use higher time steps in the GRU model. In summary, the models analyzed in this study have their advantages and disadvantages in different scenarios, and neither outperforms the others consistently across all cities. Therefore, we should select an optimized model for different tasks based on real-life scenarios, which we will discuss in Section 5.

\section{Discussion}

Predictor variable performance varied with each city as shown by the random forest results in Figures 6 \& 7 (bottom rows). The order of variable importance was similar for Dallas, Minneapolis, and Boston, while Portland's deviated from the rest, which highlights the different meteorological drivers of heat events at the various locations. Surface temperature was the most important predictor in Dallas, Minneapolis, and Boston and the second most important predictor in Portland for both the random forest's regression (predicting the value of next day heat index) and classification models (whether or not the next day heat index would be met). This result was unsurprising as temperature has an imperative role in the heat index equation and morning temperatures are positively correlated with next day temperatures. 500 hPa geopotential height was the most important predictor in Portland and the second most important predictor for the other three cities. Geopotential height relates to the overall synoptic pattern driving weather in a region. While synoptic patterns are important extreme heat drivers throughout the USA, of the four locations, Portland is most dependent upon specific synoptic patterns for extreme heat to occur. Surface dew point is another significant predictor variable and was third most important for Dallas, Minneapolis, and Boston, but in Portland 850 hPa was a better predictor than surface dew point. Low-level relative humidity can also represent cloud cover and moisture in the mixing layer, while dew point does not, which may play a role in Portland's differing dew point results. Low-level relative humidity had moderate impact in Dallas and Minneapolis and was one of the lowest impact variables in Boston.

The subsidence rate was not a strong predictor in any city and never ranked among the top five predictors. However, it generally had a moderate impact. Interestingly, 3 a.m.-9 a.m. precipitation was one of the worst predictors in Boston, Minneapolis, and Portland but had the fifth largest impact in Dallas (although it paled in comparison to the top three).

As discussed in the Data section, Portland was the only city for which the northerly wind direction issues could apply. Despite this, both 850 hPa and surface level wind directions were more influential predictor variables than in any of the three other cities. Both scored around 0.09 for the mean decrease in impurity compared to values below 0.05 for the other three locations. Due to its synoptic dependence to drive extreme heat events and complex nearby topography, Portland is especially sensitive to the direction of air flow or wind. When air flow is from the east or northeast, warm air is drawn towards Portland, whereas air flow from the south or west (typical May-September wind directions in Portland) brings cool air to Portland. Northeasterly and easterly winds are infrequent in Portland, but abundant during extreme heat periods indicating wind direction should have a heightened importance and hinting that the the northerly wind overlap issue (discussed in the Data section) may have weakened the wind direction dependence in Portland. Also note that 850 hPa wind direction outperformed surface winds, as without interference from friction at the surface, 850 wind direction better represents the true direction of flow.

Variable importance was more well-dispersed for the classification model compared to the regression model. At all four cities, the most important variable of the regression model was responsible for at least 40\% of the model based impurity, the second most important variable provided tepid values of model based impurity between 0.1 and 0.2, and all other values barely registered on the figure. Dallas was particularly extreme with a whopping 90\% due to its main variable, surface temperature. This result can be explained by Dallas's geography and climatology during the May-September period. Dallas is located in the southern plains near the Gulf of Mexico, which supplies copious amounts of moisture in the atmosphere, thus Dallas is in a perpetual state of high humidity for much of May-September. Given the high humidity, nearly every time the temperature approaches 90\degree F, the heat index will reach the 90\degree F threshold. This occurs less in a location like Portland where heat events and high humidity coincide less frequently.

Performance by model, method, and city depends on one's objective and the weighting of certain performance metrics. From a modeling perspective, overall accuracy of the model may be prioritized. However, evaluating primarily by the accuracy metric eliminates important context. From a forecasting perspective, it is more important that the model correctly predicts days of extreme heat correctly so the public can be warned. In this scenario, a false negative, classifying the next day as a non-extreme heat day when an extreme heat day occurs, has worse public ramifications than a false positive, classifying the next day as an extreme heat day when an extreme heat day does not occur (as it is worse to endanger the public than it is to issue a false alarm). From this perspective, the recall rate and precision metrics are more important. Once again, the simplified perspective leaves out context. While forecasters primarily want to avoid false positives, too many forecasts where extreme heat does not occur will negatively impact the public's confidence in the forecast, so raw recall rate is not enough to judge model performance and multiple metrics should be evaluated. Generally, there is a tradeoff between recall rate and accuracy. This relationship is seen not only by city but also by method. The GRU model (timestep = 1) outperforms both regression and classification random forest models in terms of accuracy, often by over 10\%. These accuracy improvements come at a precision cost and precision frequently drops by 20\% or more. Minneapolis is the only location where switching from random forest to GRU yields a higher precision (100\%) but this anomalous value is just a feature of variability as the number of testing days with an above 90\degree F heat index is very low (under 10) and is unlikely to be accurate with a larger dataset.

Minneapolis's model performance is an example of how models struggle to accurately predict extreme heat (or extremes in general) when little data meets the criterion. Roughly 8 \% of May-September days in Minneapolis surpass a heat index of 90\degree F. All three of the models, random forest regression, random forest classification, and GRU yield high accuracy, but these totals are artificially inflated. Since heat index values above 90\degree F are uncommon, the models are biased and barely forecast any next-day heat events. This leads to exceptional performance when most of the days do not have extreme heat, but very poor performance on days when extreme heat does occur, hence the recall rates of 21, 42, and 61\% respectively. In Figure 6 (top row), all locations show high accuracy on the training dataset and weak initial accuracy on the testing dataset. As the depth of the random forest trees increases, the testing dataset typically improves and plateaus. The testing accuracy for Boston, Portland, and Dallas all stay within 20\% of the training accuracy, but Minneapolis' testing accuracy never even reaches 50\%. In contrast, model performance in Dallas is superb due to the abundance of high heat index days. Model accuracy never drops below 82\% in Dallas, precision is typically above 80\%, and its MAE/MSE leads all four cities in all three regression model categories. F1 testing performance is also notably higher than the overall testing data accuracy for Dallas. It is possible that most southern USA cities would perform very well on the Dallas model or a similar model because of the high humidity levels in the region, but application of existing city models to other cities within their climatology region is outside the scope of our analysis.

Boston's model performance is an example of the opposite challenge as its recall values are better than those at Minneapolis and Portland, but its accuracy and MAE/MSE are the lowest of the four. Analysis of Boston's heat index distribution (Figure 2) may address this conundrum. The majority of May-September days in Boston have heat index values between 65\degree and 90\degree F with the peak of heat index values occurring just before the threshold of 80\degree F. This distribution makes it easy to predict values that are close, but slightly different in magnitude than the correct heat index value that will show up in the incorrect classification category. Generally, the models perform well on heat index values in the 80's and 90's range. However models struggle to predict extremely high heat index when values above 100\degree F which leads to slightly better classification performance than regression performance for random forest models. 

\section{Conclusion}

This study yields a myriad of results addressing several key questions including: 

    (1) Can we use machine learning models (random forest and GRU) to accurately forecast the next day heat index?
    
    (2) How do the results differ depending upon one's objective (focus on overall accuracy, or accuracy of extreme heat index days, accuracy vs precision tradeoff)?
    
    (3) What are key predictor of extreme heat events at the four locations and how do predictor variables of extreme heat vary in importance for the regression and classification models?
    
    (4) How does the performance of random forest compare to GRU and how does it depend on the performance metric?

To answer the above questions: (1) Yes, we can use machine learning models to accurately forecast next day heat index. Generally, the machine learning models applied in this study performed marginally-to-very well at predicting next day heat index values. Accuracy rates exceeded 80\% for all models at all locations other than Boston. Given a 24-hour lead time, most machine learning errors are small, between 4-6\degree F, especially relative to physics-based numerical prediction models which may exceed 10\degree F. This result suggests utility for such machine learning methods in particular given they are more cost-efficient than NWP and easily re-trainable to any location in the USA. (2) Model performance does depend on one's objective and the weighting of performance metrics. At several locations, the accuracy is higher than precision and recall, though Dallas yields very high values in all three metrics, largely due to its large sample of days meeting the 90\degree + threshold. Locations with fewer days above the threshold tend to perform poorly on accuracy and recall (except when using the GRU single timestep model). (3) At all four locations, surface temperature and geopotential height were the most critical variables for predicting the next-day heat index. This result held true for both the regression and classifier versions of the random forest models. All other variables were significantly less important in the random forest regression model. For the classifier model, however, surface dew point was important at the three more humid locations (Boston, Dallas, and Minneapolis). In Portland, a number of other variables were important in the random forest classifier model but not the random forest regression model. (4) Accuracy rates were generally higher for the GRU models across all four locations. In particular the single step GRU model exceeded the accuracy, recall, and precision of both random forest models. In contrast, MAE and MSE results between random forest and GRU were either similar (Dallas and Minneapolis) or much lower for random forest (Boston and Portlant) especially in Portland.

This study focuses on developing a lightweight, handy tool that predicts the heat index at the local level and provides insight into the issuance of extreme heat warnings. Although the project has a significant scope and includes evaluations of geographic, atmospheric, and modeling methods, there are still many ways it could be improved and explored in greater depth, from both the perspective of machine learning model accuracy and that of an improved forecasting tool. For example, using more grids of data centered at the forecast locations to create 2D or 3D graphics instead of a 1D data array could increase model accuracy using deep learning neural networks. Additionally, rather than using limited variables to predict the heat index for the next day, machine learning models could be trained with abundant data and computing power to predict HI and extreme heatwave events over a longer period of time---even beyond a couple of weeks, which is difficult for physics-based models to forecast. The use of this type of model will greatly improve the lead time for extreme heat warnings and enhance public preparedness.
{

\section{References}
\small
[1] Astsatryan, H., Grigoryan, H., Poghosyan, A., Abrahamyan, R., Asmaryan, S., Muradyan, V., ... and Giuliani, G. (2021). Air temperature forecasting using artificial neural network for Ararat valley. Earth Science Informatics, 14(2), 711-722.

[2] Benson, D.O. and Dirmeyer, P.A., 2021. Characterizing the relationship between temperature and soil moisture extremes and their role in the exacerbation of heat waves over the contiguous United States. Journal of Climate, 34(6), pp.2175-2187.

[3] Bi, K., Xie, L., Zhang, H., Chen, X., Gu, X., and Tian, Q. (2022). Pangu-weather: A 3d high-resolution model for fast and accurate global weather forecast. arXiv preprint arXiv:2211.0255

[4] Chung, J., Gulcehre, C., Cho, K., and Bengio, Y. (2014). Empirical evaluation of gated recurrent neural networks on sequence modeling. arXiv preprint arXiv:1412.3555.

[5] Clemesha, R.E., Guirguis, K., Gershunov, A., Small, I.J. and Tardy, A., 2018. California heat waves: Their spatial evolution, variation, and coastal modulation by low clouds. Climate Dynamics, 50(11), pp.4285-4301.

[6] Dahl, K., Licker, R., Abatzoglou, J.T. and Declet-Barreto, J., 2019. Increased frequency of and population exposure to extreme heat index days in the United States during the 21st century. Environmental Research Communications, 1(7), p.075002.

[7] Grotjahn, Richard, and Ghislain Faure. "Composite predictor maps of extraordinary weather events in the Sacramento, California, region." Weather and Forecasting 23.3 (2008): 313-335.

[8] Hill, A. J., G. R. Herman, and R. S. Schumacher, 2020: Forecasting Severe Weather with random forest. Mon. Wea. Rev., 148, 2135–2161, https://doi.org/10.1175/MWR-D-19-0344.1.

[9] Kantha Rao, B., and Rakesh, V. (2019). Evaluation of WRF-simulated multilevel soil moisture, 2-m air temperature, and 2-m relative humidity against in situ observations in India. Pure and Applied Geophysics, 176(4), 1807-1826.

[10] Lam, R., Sanchez-Gonzalez, A., Willson, M., Wirnsberger, P., Fortunato, M., Alet, F., Ravuri, S., Ewalds, T., Eaton-Rosen, Z., Hu, W. and Merose, A., 2023. Learning skillful medium-range global weather forecasting. Science, 382(6677), pp.1416-1421.

[11] Lian, J., Dong, P., Zhang, Y., and Pan, J. (2020). A novel deep learning approach for tropical cyclone track prediction based on auto-encoder and gated recurrent unit networks. Applied Sciences, 10(11), 3965.

[12] Perkins, S.E., Alexander, L.V. and Nairn, J.R., 2012. Increasing frequency, intensity and duration of observed global heatwaves and warm spells. Geophysical Research Letters, 39(20).

[13] Rothfusz, L. P. (1990). The heat index equation (or, more than you ever wanted to know about heat index). Fort Worth, Texas: National Oceanic and Atmospheric Administration, National Weather Service, Office of Meteorology, 9023, 640.

[14] Schultz, M. G., Betancourt, C., Gong, B., Kleinert, F., Langguth, M., Leufen, L. H., ... and Stadtler, S. (2021). Can deep learning beat numerical weather prediction?. Philosophical Transactions of the Royal Society A: Mathematical, Physical and Engineering Sciences, 379(2194).

[15] Steadman, R.G., 1979: The assessment of sultriness. Part I: A temperature-humidity index based on
human physiology and clothing science. J. Appl. Meteor., 18, 861-873

[16] U.S. EPA (U.S. Environmental Protection Agency). 2022. Climate Change Indicators: Heat Waves https://www.epa.gov/climate-indicators/climate-change-indicators-heat-waves

[17] Yu, Y., Chen, S. H., Huang, C. C., Paw U, K. T., Schmitt, C., Zhao, Z., ... and Lo, M. H. (2025). A numerical study of the agricultural irrigation effects on summer soil moisture and near-surface meteorology in California’s Central Valley. Journal of Hydrometeorology, 26(6), 641-659.

[18] Zhang, D., and Kabuka, M. R. (2018). Combining weather condition data to predict traffic flow: a GRU‐based deep learning approach. IET Intelligent Transport Systems, 12(7), 578-585.
}
\end{document}